\documentclass[prb,english,graphics,showpacs]{revtex4}
\usepackage{graphicx}
\usepackage{amssymb,amsmath,amsfonts,mathrsfs,eufrak,fancyhdr}
\usepackage{theorem,eepic}

\def\dim{\mathsf{dim}}

\def\qed{$\,\blacksquare$\par\bigskip}
\def\vec#1{{\boldsymbol{#1}}}
\def\<{\langle}
\def\>{\rangle}

\def\set#1{{\sf #1}}
\def\alg#1{{\mathfrak #1}}
\def\map#1{{\mathcal{#1}}}
\def\hil#1{{\mathscr{#1}}}

\def\span{\set{span}}

\def\Cmplx{\mathbb C}
\def\Naturals{\mathbb N}
\def\Intgrs{\mathbb Z}

\def\Tr{\operatorname{Tr}}

\newtheorem{lemma}{Lemma}[section]
\newtheorem{proposition}[lemma]{Proposition}
\newtheorem{corollary}[lemma]{Corollary}

\newtheorem{theorem}[lemma]{Theorem}

\begin{document}
\title{Asymptotic evolution of Random Unitary Operations}
\author{J. Novotn\'y$^{(1,2)}$, G. Alber$^{(2)}$, I. Jex$^{(1)}$}
\affiliation{$^{(1)}$~Department of Physics, FJFI \v CVUT v Praze, B\v
rehov\'a 7,
115 19 Praha 1 - Star\'e M\v{e}sto, Czech Republic\\
$^{(2)}$~Institut f\"ur Angewandte Physik, Technische
Universit\"at Darmstadt, D-64289 Darmstadt, Germany}
\date{\today}

\begin{abstract} We analyze the asymptotic dynamics
of quantum systems resulting from large numbers of iterations of random unitary operations.
Although, in general, these quantum operations cannot be diagonalized it is shown that their resulting
asymptotic dynamics is described by a diagonalizable superoperator.
We prove that this asymptotic dynamics takes place in a typically low dimensional attractor space
which
is independent of the probability distribution of the unitary operations applied.
This vector space is
spanned by all eigenvectors of the
unitary operations involved which are associated with eigenvalues of unit modulus.
Implications for possible asymptotic dynamics of iterated random unitary operations are presented and exemplified in
an example involving random
controlled-not operations acting on two qubits.
\end{abstract}

\maketitle

\section{Introduction}
In recent years the rapid advancement of
quantum technology with its capabilities
of controlling
individual quantum systems has given
rise to impressive developments
in the areas of quantum information science and high-precision
quantum metrology.\cite{individual}  In particular, current experiments on large ensembles of
interacting quantum
systems open interesting perspectives
to investigate in detail not
only the transition from quantum to classical behavior but also to trace down
those quantum phenomena or effects that still are observable on the
mesoscopic or macroscopic scale. A paradigm
of such large physical systems are
interacting networks whose dynamics is currently
investigated  intensively in the
classical domain.\cite{Newton2003} Such networks are capable of simulating
the behavior of real world systems like the internet or social
dynamics.\cite{Barabasi} Typically, in these systems a number of modes
representing
physical objects are coupled to each other by random
interactions. A particularly  interesting issue is to determine the
dynamics of the system. In view of these current activities the
natural question arises which characteristic properties govern the dynamics of such networks if
each classical node is replaced by a quantum system
and, correspondingly, the classical interactions by quantum operations.

In general,
determining the time evolution of large quantum systems is
difficult and analytic or closed-form solutions are possible in exceptional
cases only. In particular, this applies to the dynamics of open quantum systems
in which a large quantum
system is in contact with an additional physical
system. The influence of such an external system can be taken into
account in various ways. In special cases it may be described
by randomly applied unitary operations.
Such a case is realized, for example, if
the nodes of a large quantum network represent participants of a quantum
communication network and if these nodes
establish node-to-node communication in a random way by using
quantum  protocols
which can be described by unitary transformations.
A natural question arising in this context is
what is the resulting quantum state of the network after a large number of
such communication steps.
More generally, such a quantum network involving random unitary transformations
can characterize the dynamics of any interacting quantum
system in which
the interactions involved can be described by repeatedly applied random
unitary transformations.

A natural approach to determine the dynamics of a quantum system involves
diagonalization of the generator of the time evolution. This way the dynamics
can be determined in a convenient way even in the asymptotic limit of
arbitrarily long interaction times.
The situation becomes significantly
more complicated for open quantum systems because the relevant
generators are often
non-hermitian and not normal \cite{linear_algebra}
so that they cannot be diagonalized.
Nevertheless, in such cases it is still possible to use the Jordan
canonical form (see Appendix \ref{app:jordan}) of these operators for determining the dynamics for arbitrarily long interaction times.
This leads to the highly nontrivial problem of handling generalized
eigenvectors of the relevant generators which are in general not orthogonal.

Motivated by these aspects
in this paper we address the problem of determining general properties
of the asymptotic dynamics
of quantum systems whose dynamics is governed by repeated applications
of random unitary transformations.
A main goal of this paper is to demonstrate that the Jordan canonical form of
the generators of
random unitary transformations
have rather unexpected and useful special properties which
allow to obtain even
closed-form expressions for the asymptotic quantum state
resulting from a large number of iterations of random unitary transformations.
It will be proved that
there is always a vector subspace of so-called attractors on
which the resulting superoperator governing the iterative time evolution
of quantum states
can be diagonalized and in which the asymptotic quantum dynamics
takes place. As a main result a structure
theorem is derived for this set of attractors which allows to determine them
in a convenient way. Furthermore, it is shown how
the asymptotic iterative dynamics of arbitrary quantum states
can be written in terms of these attractors.
Based on these findings we
show that in general the asymptotic dynamics is non-monotonic.
Finally,
aspects of these general properties
are exemplified by studying in detail the dynamics of
two qubits which are coupled by randomly applied controlled-not operations.
It should be mentioned that some of the results characterizing the asymptotic dynamics
can also be obtained by a different
approach which uses special properties of random unitary transformations in order to construct
a convenient Ljapunov function.\cite{NAJ1}

This paper is structured as follows. In Section
\ref{section_preliminaries} we summarize basic properties
of random unitary transformations which are useful for our subsequent
discussion.
In Section
\ref{section_jordan_cannonical_form} we examine special properties
of the Jordan canonical form of random unitary maps. The central
statement of the paper, namely the structure theorem for attractors of
random unitary operations,
is derived in Section \ref{section_structure theorem_for
attractors}.  Characteristic properties of
attractors are investigated in Section \ref{section_properties_of
attractors}. Section \ref{sec_discussion} is devoted to important
implications resulting from the structure theorem. Finally, as an example the
asymptotic dynamics of two qubits which are coupled by random controlled-not operations
is discussed on the basis of our general results in  Section \ref{section_cnots}.

\section{Basic properties of random unitary operations}\label{section_preliminaries}
A random unitary operation (RUO) $\Phi$ is a completely
positive trace-preserving map admitting a convex decomposition of
the form \cite{Holevo2001}
\begin{equation}
\label{unitary_channel}  \Phi(\rho)= \sum_{i=1}^m p_{i} U_{i} \rho
U_{i}^{\dagger}.
\end{equation}
Thereby, $U_i$ denotes a unitary
operator acting on a Hilbert space $\hil H$ and this unitary operation is applied
onto the quantum state
$\rho$ with probability $p_i > 0$ so that $\sum_{i=1}^m p_i= 1$.
These latter probabilities take into account
classical uncertainties in the realizations of the unitary quantum
evolution involved. This uncertainty can be the result of an unknown error mechanism or
of an unknown unitary evolution involving an additional ancillary system.
In the following
we are interested in the asymptotic dynamics resulting from many iterative applications of $\Phi$.
Starting with our quantum system in the initial state $\rho(0)$, the
$(n+1)$-st step of this iteration procedure changes the state
after the $n$-th iteration $\rho(n)$ to the state $\rho(n+1)=\Phi(\rho(n))$. Our aim is to
analyze the asymptotic behaviour of this iteration procedure.
The random unitary map $\Phi$ of Eq.(\ref{unitary_channel}) belongs to
the class of bistochastic or doubly stochastic maps
\cite{Bengtsson2006,Bhatia2007} which leave the maximally mixed
state invariant, i.e.
\begin{equation}
\Phi(I)=\sum_{i=1}^m p_i U_i U_i^{\dagger} = I,
\end{equation}
and it
acts on the Hilbert space $\map B(\hil H)$ of all linear operators defined on a d-dimensional Hilbert space $\hil H$.
The dimension of the input and output system is the same.
The Hilbert space $\map B(\hil H)$ is equipped with the Hilbert-Schmidt inner product
$(A,B)_{HS} = \Tr(A^{\dagger}B)$ for all $A,B \in \mathcal{B}(H)$.
The adjoint operator of $\Phi$ is given by
\begin{equation}
\Phi^{\dagger}(A)=\sum_{i=1}^{m} p_{i} U_{i}^{\dagger} A U_{i}.
\end{equation}

In general, the RUO $\Phi$ is neither hermitian nor normal and
consequently is not diagonalizable. Therefore, its resulting iterated dynamics
has to be analyzed with the help of Jordan normal forms
\cite{linear_algebra} (see Appendix \ref{app:jordan}).
It is a main goal of our subsequent discussion to prove that the Jordan normal forms of RUOs have
interesting special properties which are particularly useful for the description of their asymptotic
iterated
dynamics.
In particular, there exists a Jordan base in the Hilbert space
$\mathcal{B}(\hil H)$ in which the matrix of the map (\ref{unitary_channel}) has
a block diagonal form (\ref{jordan_form}).

Let us formulate first several characteristic properties of RUOs.
\begin{lemma}
The random unitary map $\Phi$ defined by the relation
(\ref{unitary_channel}) fulfills the following properties:
\begin{itemize}
\item[1)] The norm of the RUO $\Phi$ induced by the
Hilbert-Schmidt norm of the Hilbert space $\map B(\hil H)$ equals
unity. \item[2)] If $\lambda$ is an eigenvalue of the map $\Phi$,
then $|\lambda| \leq 1$. \item[3)] Let $X_{\lambda} \in
\mathcal{B}(\hil H)$ be a generalized eigenvector corresponding to
the eigenvalue $\lambda$ of the map $\Phi$, then $\lambda = 1$ or
$\Tr{X_{\lambda}}=0$.
\end{itemize}
\end{lemma}
{\bf Proofs.}\\
(1-2) First we prove that the Hilbert-Schmidt norm is
unitarily invariant. For this purpose consider an arbitrary operator $A \in
\mathcal{B}(\hil H)$ and two unitary operators $U,V \in
\mathcal{B}(\hil H)$. As a trace of matrix products
is invariant under cyclic permutations we get
\begin{equation}
||UAV||_{HS}=\left\{\Tr\left[(UAV)^{\dagger}(UAV)\right]\right\}^{\frac{1}{2}}
=\left\{\Tr(A^{\dagger}A)\right\}^{\frac{1}{2}}=||A||_{HS}.
\end{equation}
Therefore one can show that $||\Phi||=\sup_{||A||_{HS}\leq 1}
||\Phi(A)||_{HS}=1 $. Let $A \in \mathcal{B}(\hil H)$, then the
Hilbert-Schmidt norm of the operator $\Phi(A)$ is bounded by
\begin{equation}
||\Phi(A)||_{HS} = \left| \left| \sum_{i} p_{i} U_{i}AU_{i}^{\dagger}
\right|\right|_{HS} \leq \sum_{i} p_{i} \left| \left|
U_{i}AU_{i}^{\dagger} \right|\right|_{HS} = \left| \left| A
\right|\right|_{HS}.
\end{equation}
Moreover, we have $||\Phi(I)||_{HS}=||I||_{HS}$. Hence, $\left|
\left|
\Phi \right|\right| = 1$ and consequently $|\lambda| \leq 1$.\\
(3) If $X_{\lambda}$ is a generalized eigenvector corresponding to an
eigenvalue $\lambda$ of the map $\Phi$, then there is a $n \in \Naturals$
such that $(\Phi - \lambda I)^{n} (X_{\lambda}) = 0$ because a simple
calculation yields
\begin{equation}
\Tr \left\{(\Phi - \lambda I)^{n} (X_{\lambda})\right\} =
(1-\lambda)^n \Tr X_{\lambda} = 0.
\end{equation}
This equation
can be fulfilled only if $\lambda = 1$ or $\Tr{X_{\lambda}}=0$.
\\
\rightline{\qed}

Thus, all Jordan blocks in the Jordan normal decomposition of the map
$\Phi$ correspond to eigenvalues $\lambda$ with $|\lambda| \leq 1$.
For our subsequent discussion let us introduce the following notation. Suppose
that $\lambda$ is an eigenvalue of the map $\Phi$. We denote the
corresponding eigen-subspace by
$\set{Ker}(\Phi- \lambda I)$, i.e.
\begin{equation}
\set{Ker}(\Phi- \lambda I)=\left\{X \in \mathcal{B}(\hil H)|
\Phi(X)=\lambda X\right\} ,
\end{equation}
and the range of the map $\Phi-\lambda I$ by $\set{Ran}(\Phi-
\lambda I)$, i.e.
\begin{equation}
\set{Ran}(\Phi- \lambda I)=\left\{X \in \mathcal{B}(\hil H)|
\exists Y \in \mathcal{B}(\hil H), \hspace{0.5em}
X=\Phi(Y)-\lambda Y\right\}.
\end{equation}
Furthermore,
let us define $d_{\lambda}=\dim{(\set{Ker}(\Phi- \lambda I))}$ and $\sigma_{|1|}$ as the set of all eigenvalues of the
linear map $\Phi$ satisfying $|\lambda|=1$. Finally, the vector
subspace spanned by all eigenstates corresponding to eigenvalues
$\lambda$ with $|\lambda| \leq 1$
we call \emph{the attractor space of the RUO} $\Phi$ and
denote it by $\set{Atr}(\Phi)$, i.e.
\begin{equation}
\set{Atr}(\Phi) = \bigoplus_{\lambda \in
\sigma_{|1|}}\set{Ker}(\Phi- \lambda I).
\end{equation}
We call elements of this subspace attractors of the dynamics because, as we
will show later, the asymptotic iterated dynamics of the RUO is
completely determined by these linear operators.

\section{Jordan canonical form of Random unitary operations}
\label{section_jordan_cannonical_form}
In this section
we prove that all
Jordan blocks corresponding to eigenvalues
$\lambda$ with $|\lambda| \leq 1$
are one-dimensional. In other words, generalized eigenvectors
corresponding to eigenvalues $|\lambda|=1$ are all eigenvectors.
This statement is equivalent to the following theorem (for details see Appendix \ref{app:jordan}).
\begin{theorem}
\label{theorem_diagonalozation} Let $\Phi:\mathcal{B}(\hil H)
\rightarrow \mathcal{B}(\hil H)$ be a random unitary operation
defined by (\ref{unitary_channel}) and $\lambda$ its eigenvalue
satisfying $|\lambda|=1$, then we have
\begin{equation}
\set{Ker}(\Phi-\lambda I) \cap \set{Ran}(\Phi-\lambda I)=\{0\}.
\end{equation}
\end{theorem}
{\bf Proof.} We prove this theorem by contradiction. Suppose
there is an operator $0 \neq A\in \mathcal{B}(\hil H)$ and $A
\in \set{Ker}(\Phi-\lambda I) \cap \set{Ran}(\Phi-\lambda I)$. This implies
$\Phi(A)=\lambda A$ and there is an operator $0 \neq B \in
\mathcal{B}(\hil H)$ such that $\Phi(B)= \lambda B+A$. By
induction one can conclude
\begin{equation}
\Phi^n(B)=\lambda^n B + n\lambda^{n-1}A
\end{equation}
and consequently
\begin{equation}
n\left|\left|A \right|\right|-\left|\left|B\right|\right| \leq
\left|\left| \lambda^n B + n\lambda^{n-1}A \right|\right| =
\left|\left| \Phi^n(B) \right|\right| \leq \left|\left|\Phi
\right|\right|^n \left|\left| B \right|\right| = \left|\left| B
\right|\right|.
\end{equation}
Because the resulting inequality
\begin{equation}
\left|\left|A\right|\right| \leq
\frac{2}{n}\left|\left|B\right|\right|
\end{equation}
has to be fulfilled for arbitrary $n \in \Naturals$ the only
alternative left is that $A=0$.
\\ \rightline{\qed}

Let $Y_{j,k}$ ($j \in \hat{p}$, $k \in \left\{1,2,
\hdots,\dim(J_j)\right\}$) (compare with Appendix \ref{app:jordan}) be the Jordan basis of the RUO $\Phi$.
$J_j$ is a Jordan block corresponding to an eigenvalue $\lambda_j$
with a basis formed by the generalized eigenvectors $Y_{j,k}$ ($k
\in \left\{1,2, \hdots,\dim(J_j)\right\}$). Let $\rho(0) \in
\mathcal{B}(\hil H)$ be an input density operator. We denote by
$\beta_{j,k}^{(0)}$ the parameters of the unique decomposition of
the density operator $\rho(0) \in \mathcal{B}(\hil H)$ into this
basis, i.e.
\begin{equation}
\label{decomposition} \rho(0)=
\sum_{j=1}^{p}\sum_{k=1}^{\dim(J_j)} \beta_{j,k}^{(0)}Y_{j,k}.
\end{equation}
Consider now the density operator $\rho(n)=\Phi^n(\rho(0))$ describing the
physical system after $n$ iterations and
denote its decomposition coefficients (\ref{decomposition})
into the same basis by $\beta_{j,k}^{(n)}$. It is clear that
the coefficients $\beta_{j,k}^{(n)}$ corresponding to eigenvectors of
eigenvalues $\lambda_{j} \in \sigma_{|1|}$ evolve simply as
\begin{equation}
\label{fixed_evolution} \beta_{j}^{(n)}=\lambda_j^n
\beta_{j}^{(0)}.
\end{equation}
(We omit the second index $k$ intentionally because in this case all
the Jordan blocks are one dimensional.)

Now we have to analyze the behavior
of the remaining coefficients. It is governed by
the following theorem which quantifies how the remaining coefficients
$\beta_{j,k}^{(n)}$, corresponding to Jordan vectors $Y_{j,k}$ with
$|\lambda_j| < 1$, evolve.
\begin{theorem}
\label{theorem:vanishing part} Let $\Phi: \mathcal{B}(\hil H)
\rightarrow \mathcal{B}(\hil H)$ be a quantum random unitary
operation defined by (\ref{unitary_channel}) with its Jordan basis
$Y_{j,k}$ ($j \in \hat{p}$, $k \in \left\{1,2,
\hdots,\dim(J_j)\right\}$) and $\rho(0) \in \mathcal{B}(\hil H)$
be an input density operator.
Furthermore,
let $\beta_{j,k}^{(n)}$ be the decomposition coefficients of
$\rho(n)=\Phi^n(\rho(0))$ into this Jordan basis, i.e.
\begin{equation}
\label{decomposition_n} \rho(n)=
\sum_{j=1}^{p}\sum_{k=1}^{\dim(J_j)} \beta_{j,k}^{(n)}Y_{j,k}.
\end{equation}
For any
eigenvalue $\lambda_s$ ($|\lambda_s| <1$) of the map
$\Phi$ with its corresponding Jordan block $J_s$ and its Jordan
chain $Y_{s,k}$ ($k \in \left\{1,2, \hdots,\dim(J_s)\right\}$)
the coefficients $\beta_{s,k}^{(n)}$ vanish in the limit of large
$n$
\begin{equation}
\lim_{n \rightarrow +\infty} \beta_{s,k}^{(n)} \rightarrow 0,
\hspace{2em} {\rm for} \hspace{1em} \forall k \in \left\{1,2,
\hdots,\dim(J_s)\right\}.
\end{equation}
\end{theorem}
{\bf Proof.} This theorem follows directly from the fact that the Jordan block $(J_s)^n$ of dimension $\dim{(J_s)}$
with
\begin{equation}
J_{s}= \left( \begin{array}{cccc} \lambda_s & 1& & \\ & \lambda_s &\ddots & \\
& & \ddots & 1\\ & & & \lambda_s
\end{array}\right)
\end{equation}
vanishes in the limit of large numbers of iterations $n$, i.e.
\begin{equation}
\lim_{n \rightarrow \infty} (J_s)^n = 0.
\end{equation}
One can check that the entry $(J_s^n)_{ij}$ ($i \leq j \leq \dim{J_s}$)
of the upper triangular matrix $(J_s)^n$ fulfills the inequality
\begin{equation}
|(J_s^n)_{ij}|=|\lambda_s|^{n-(j-i)}
{n\choose n-(j-i)}\leq |\lambda|^{n-\dim{J_s}} n^{\dim{J_s}}
\end{equation}
so that we obtain the relation
\begin{equation}
\lim_{n \rightarrow \infty} (J_s^n)_{ij} = 0, \hspace{2em} \forall i,j \in \{ 1, \ldots, \dim{J_s}\}.
\end{equation}
\\ \rightline{\qed}

In view of this theorem
the asymptotic dynamics of the state $\rho(0)$
under iterations of the random unitary operation $\Phi$ is given completely in terms of
its attractors. The remaining coefficients of the decomposition
of the initial state $\rho(0)$ (\ref{decomposition_n}) become vanishingly small after
sufficiently many iterations of the map. An interesting question which will be addressed in the following
is how to determine the set of attractors.

\section{Structure theorem for attractors}
\label{section_structure theorem_for attractors} Let us now
study the structure of the attractors, i.e. of
all eigenspaces $\set{Ker}(\Phi-\lambda
I)$, with
$\lambda \in \sigma_{|1|}$. In the case of
random unitary operations the
following powerful theorem can be proved
which allows us to specify the space of attractors of the RUO $\Phi$.
In this context
it should be also mentioned that for the more general case of arbitrary unital quantum operations interesting
general results have been derived by Kribs \cite{Kribs2003,Kribs2003b} recently.
\begin{theorem}
\label{comutator} Let $\Phi:\mathcal{B}(\hil H) \rightarrow
\mathcal{B}(\hil H)$ be a random unitary map $\Phi$
(\ref{unitary_channel}) and $\lambda \in \sigma_{|1|}$. Then the
eigenspace $\set{Ker}(\Phi-\lambda I)$ corresponding to this
eigenvalue $\lambda$ is equal to the set
 \begin{equation}
D_{\lambda} := \left\{X \in
\mathcal{B}(\hil H)|U_iX=\lambda XU_i \hspace{0.5em} {\rm for}
 \hspace{0.5 em} i=1, \hdots, m
\right\}.
 \end{equation}
\end{theorem}
{\bf Proof.} The map $\Phi$ is unital, that is $\Phi(I)=I$.
Therefore, every $X \in D_{\lambda}$ fulfils $\Phi(X)=\lambda X$ and
thus $D_{\lambda} \subset \set{Ker}(\Phi-\lambda I)$. To prove the
converse, let us consider $X \in \set{Ker}(\Phi-\lambda I)$. If $X=0$, then $X \in
D_{\lambda}$. So let us assume that $X \neq 0$. Using the unitary invariance
of the Hilbert-Schmidt norm we get
\begin{equation}
\label{inequality_equality} \left|\left|X \right|\right| =
\left|\left|\lambda X \right|\right|=\left|\left|\sum_{i=1}^{m}
p_i U_i X U_i^{\dagger} \right|\right| \leq \sum_{i=1}^{m}p_i
\left|\left|U_i X U_i^{\dagger}\right|\right| =
\left|\left|X\right|\right|.
\end{equation}
Therefore, the inequality (\ref{inequality_equality}) is in fact an equality and can be rewritten in the form
\begin{equation}
\left( \sum_{i=1}^{m}p_i v_i,\sum_{i=1}^{m}p_i
v_i\right)=\left(\sum_{i=1}^{m}p_i
(v_i,v_i)^{\frac{1}{2}}\right)^2
\end{equation}
with $v_i=U_iXU_i^{\dagger}$. Hence we get
\begin{eqnarray}
\label{inequality1}
\sum_{i<j}2p_ip_j
(v_i,v_i)^{\frac{1}{2}}(v_j,v_j)^{\frac{1}{2}}&=&\sum_{i<j}p_ip_j\left[(v_i,v_j)+(v_j,v_i)\right]=
\sum_{i<j}2p_ip_j {\rm Re}(v_i,v_j) \leq  \sum_{i<j}2p_ip_j\left|{\rm Re}(v_i,v_j)\right|\nonumber \\
&\leq&  \sum_{i<j}2p_ip_j \left|(v_i,v_j)\right| \leq
\sum_{i<j}2p_ip_j (v_i,v_i)^{\frac{1}{2}}(v_j,v_j)^{\frac{1}{2}}.
\end{eqnarray}
Because the left and right side of the relation (\ref{inequality1}) are the same, all inequalities are actually equalities. In particular, we have
\begin{equation}
{\rm
Re}(v_i,v_j)=\left|(v_i,v_j)\right|=(v_i,v_i)^{\frac{1}{2}}(v_j,v_j)^{\frac{1}{2}}\neq
0 \hspace{1em} {\rm for \hspace{0.5em} all\hspace{0.5em}} i,j\in
\left\{1,\hdots,m\right\}
\end{equation}
which can be fulfilled if and only if $v_j=\beta_{ij} v_i$ (for all $i$,$j$) with $\beta_{ij}>0$. From the unitary invariance of the Hilbert-Schmidt norm
\begin{equation}
\left|\left|X\right|\right|=\left|\left|
U_iXU_i^{\dagger}\right|\right|=\beta_{ij}\left|\left|U_jXU_j^{\dagger}\right|\right|=\left|\left|X\right|\right|
\end{equation}
we conclude that $\beta_{ij}=1$ for all $i,j\in \{1,...,m\}$  and hence
\begin{equation}
U_1XU_1^{\dagger}=U_2XU_2^{\dagger}= \hdots = U_mXU_m^{\dagger}.
\end{equation}
Finally, using the equality $\Phi(X)=\lambda X$ we obtain
$U_iXU_i^{\dagger}=\lambda X$, i.e. $X \in D_{\lambda}$.
\\ \rightline{\qed}

As a consequence of this structure theorem \ref{comutator} the following
corollary can be proved.
\begin{corollary}
\label{theorem_orthogonality}
The random unitary operation $\Phi$ defined by
(\ref{unitary_channel}) fulfills the following properties:
\begin{itemize}
\item[1)]
If $\lambda$ is an eigenvalue of the operation $\Phi$ fulfilling
$|\lambda|=1$, then
\begin{equation}
\set{Ker}(\Phi-\lambda I)\hspace{0.3em} \bot
\hspace{0.3em}\set{Ran}(\Phi-\lambda I).
\end{equation}
\item[2)]
If $\lambda_1$,$\lambda_2$ are two different eigenvalues of the
operation $\Phi$ fulfilling $|\lambda_1|=|\lambda_2|=1$, then
\begin{equation}
\set{Ker}(\Phi-\lambda_1 I)\hspace{0.3em} \bot \hspace{0.3em}\set{Ker}(\Phi-\lambda_2 I).
\end{equation}
\end{itemize}
\end{corollary}
{\bf Proof.} First, from the theorem (\ref{comutator}) follows that if $\Phi(X) =\lambda X$ and $|\lambda| = 1$,
then $\Phi^{\dagger}(X)=\lambda^*X$ and $\Phi^{\dagger}(X^{\dagger})=\lambda X^{\dagger}$.
In order to show that the set $\set{Ker}(\Phi-\lambda I)$ is
orthogonal to the set $\set{Ran}(\Phi- \lambda I)$ we have to
prove that $(K,R)=0$ is fulfilled for arbitrary elements
$K \in \set{Ker}(\Phi-\lambda I)$ and $R \in \set{Ran}(\Phi-\lambda I)$. Therefore, there is an operator $Q \in
\mathcal{B}(\hil H)$ with $R=\sum_i p_{i}
U_{i}QU_{i}^{\dagger} - \lambda Q$. Hence, using theorem
\ref{comutator} we have the orthogonality relation
\begin{eqnarray}
(K,R)=\Tr\{K^{\dagger}R\}=\sum_{i}p_{i}\Tr\{K^{\dagger}U_{i}QU_{i}^{\dagger}\}-\lambda\Tr\{K^{\dagger}Q\}=
\lambda\sum_{i}p_{i}\Tr\{K^{\dagger}Q\}- \lambda\Tr\{K^{\dagger}Q\}=0.
\end{eqnarray}
\noindent The second property is a
consequence of the identity
\begin{equation}
\left(X_1,X_2\right)=\frac{1}{\lambda_2}\left(X_1,\Phi(X_2)\right)=\frac{1}{\lambda_2}\left(\Phi^{\dagger}
(X_1),X_2\right)=\frac{\lambda_1}{\lambda_2}\left(X_1,X_2\right)
\end{equation}
which is valid for
$X_1 \in
\set{Ker}(\Phi-\lambda_1 I)$ and $X_2 \in \set{Ker}(\Phi-\lambda_2
I)$ and for any mutually different non-zero eigenvalues
$\lambda_1$ and $\lambda_2$.
Therefore, the last equality can be satisfied only if
$\left(X_1,X_2\right)=0$.
\\ \rightline{\qed}

This corollary together with theorem
\ref{theorem_diagonalozation} has the following important consequence.
\begin{theorem}
\label{theorem:final state} Let $\Phi: \mathcal{B}(\hil H)
\rightarrow \mathcal{B}(\hil H)$ be a quantum random unitary
operation defined by (\ref{unitary_channel}) and $\rho(0) \in
\mathcal{B}(\hil H)$ be an input density operator,
then the asymptotic iterative dynamics of the state
$\rho(0)$ under the evolution map $\Phi$ is given by
\begin{equation}
\label{asymptotic_dynamics}
\rho_{\infty}(n)  = \sum_{\lambda \in
\sigma_{|1|}, j=1}^{d_{\lambda}}
\lambda^n\Tr\{\rho(0)X_{\lambda,i}^{\dagger}\} X_{\lambda,i}
\end{equation}
and satisfies the relation
\begin{equation}
\lim_{n \rightarrow \infty} || \rho(n)-\rho_{\infty}(n)|| =0
\end{equation}
with
$\rho(n)=\Phi^n(\rho(0))$ and
with the
complete set of orthonormal basis elements
$X_{\lambda,i}$ ($i \in \left\{1,2, \hdots,
d_{\lambda}\right\}$)
of
the space $\set{Ker}(\Phi-\lambda I)$.
\end{theorem}
{\bf Proof.} In order to prove this theorem we have to show that the mutually orthogonal subspaces
\begin{equation}
\alg I_0= \bigoplus_{\lambda \in \sigma_{|1|}}
\set{Ker}(\Phi-\lambda I) \hspace{1em} {\rm and} \hspace{1em} \alg
I_1= \bigcap_{\lambda \in
\sigma_{|1|}} \set{Ran}(\Phi-\lambda I)
\end{equation}
are invariant under the map $\Phi$ and that they satisfy the relation $\alg I_0 \oplus \alg I_1 = {\cal B}(\mathcal{\hil H})$.
The second claim is a direct consequence of corollary \ref{theorem_orthogonality}. The first claim follows from the fact that all subspaces $\set{Ker}(\Phi-\lambda I)$ and $\set{Ran}(\Phi-\lambda I)$ are
invariant under the map $\Phi$; that is, $\Phi(\set{Ker}(\Phi-\lambda
I))\subset \set{Ker}(\Phi-\lambda I)$ and
$\Phi(\set{Ran}(\Phi-\lambda I)) \subset \set{Ran}(\Phi-\lambda
I)$. Now we can choose some orthogonal basis vectors $X_{\lambda,i}$ in the subspaces
$\set{Ker}(\Phi-\lambda I)$ with $|\lambda«|=1$ and the Jordan basis
$Y_{j,k}$ of the map $\Phi$ restricted to the subspace $\alg I_1$.
These vectors form a basis of the Hilbert
space $\mathcal{B}(\hil H)$. Now we consider a decomposition of $\rho(0)$
into these basis vectors. As was shown in theorem \ref{theorem:vanishing
part}, the part corresponding to the subspace $\alg I_1$ vanishes
for $n \rightarrow +\infty$ and the dynamics of the state
$\rho(0)$ on the subspace $\alg I_0$ is given by
(\ref{fixed_evolution}).
\\ \rightline{\qed}
\section{Basic properties of attractors}
\label{section_properties_of attractors}
In this section we discuss some basic properties of RUOs
which are useful for obtaining
the complete set of attractors.

A basic property arises straightforwardly from the
theorem \ref{comutator}.
\begin{proposition}
\label{theorem_product of attractors}
\begin{itemize}
\item[1)] Let $X_{\lambda_1}$ and $X_{\lambda_2}$ be attractors of the RUO (\ref{unitary_channel})
corresponding to eigenvalue $\lambda_1$ and $\lambda_2$, respectively, then the product of these attractors
$X_{\lambda_1}X_{\lambda_2}$ is either an attractor corresponding to eigenvalue $\lambda_1 \lambda_2$ or it is the
zero operator.
\item[2)] Let $X_{\lambda}$ be an attractor of the RUO (\ref{unitary_channel}) corresponding to
the eigenvalue $\lambda$, then $X_{\lambda}^{\dagger}$ is also an
attractor of the RUO (\ref{unitary_channel}) corresponding to
the eigenvalue $\lambda^{*}$.
\end{itemize}
\end{proposition}
\begin{Proof}
This proposition follows from the identities
\begin{equation}
U_iX_{\lambda_1}X_{\lambda_2}=\lambda_1X_{\lambda_1}U_iX_{\lambda_2}=\lambda_1\lambda_2X_{\lambda_1}X_{\lambda_2}U_i
\end{equation}
and
\begin{equation}
U_iX_{\lambda}^{\dagger}=(X_{\lambda}U_i^{\dagger})^{\dagger}=(\lambda U_i^{\dagger}X_{\lambda})^{\dagger}=\lambda^*X_{\lambda}^{\dagger}U_i,
\end{equation}
which are valid for all $i \in \hat{m} :=\{1,\ldots,m\}$.
\\ \rightline{\qed}
\end{Proof}
Based on our preceding analysis a single step of the asymptotic dynamics is described by the superoperator
\begin{equation}
\label{assymptotic_step}
\Phi_{ass}(.) = \sum_{\lambda \in
\sigma_{|1|}, i=1}^{d_{\lambda}} \lambda \Tr\left(X_{\lambda,i}^{\dagger}(.)  \right)X_{\lambda,i},
\end{equation}
which fulfils the property
\begin{equation}
\lim_{n \rightarrow \infty}||\Phi^n - \Phi_{ass}^n||=0.
\label{Asymp1}
\end{equation}
The superoperator (\ref{assymptotic_step}) is a unital quantum operation.
In order to prove this statement, let us define the projector
\begin{equation}
\map P (.)= \sum_{\lambda \in
\sigma_{|1|}, i=1}^{d_{\lambda}} \Tr\left(X_{\lambda,i}^{\dagger}(.)  \right)X_{\lambda,i},
\end{equation}
which projects all elements of the vector space $\map B(\hil H)$ onto the attractor space $\set{Atr}(\Phi)$.
The structure theorem \ref{comutator} and the orthogonality of all elements of the attractor space ensure that
$\left[\Phi,\map P \right]=0$ and $\left[\Phi_{ass},\map P \right]=0 $. These commutation properties imply that for any integer $n$
the action of
the superoperator $\Phi^n_{ass}$ on an arbitrary operator $A \in \map B(\hil H)$ is given by
\begin{equation}
\Phi^n_{ass}(A) = \Phi^n_{ass}\left(\map P(A)\right)=U^n_{i_0}\map P(A) U_{i_0}^{\dagger n}
\end{equation}
for an arbitrary $i_0 \in \hat{m}$.
Thus, for any integer $n$ the action of the map $\Phi^n_{ass}$ on the Hilbert space ${\cal B}(\hil H)$ is a sequence of a projection onto
the attractor space and a unitary operation. As a consequence it is a completely positive map and in view of Eq.(\ref{Asymp1})
it describes the dynamics of the iterated random unitary operation in the asymptotic limit of large numbers $n$ of iterations.

It is instructive to
analyze this property of complete positivity also from another perspective by using
the concept of dynamical matrices (compare with Appendix
\ref{appendix_dynamical_matrices}).\cite{Bengtsson2006} In order to obtain the
dynamical matrix of the asymptotic map $\Phi_{ass}$ we
first calculate its matrix elements
in an orthonormal basis, i.e.
\begin{eqnarray}
\left(\Phi_{ass}\right)^{m \mu}_{n \nu}&=& \<m \mu|\Phi_{ass}|n\nu
\>= \sum_{\lambda \in \sigma_{|1|}, i=1}^{d_{\lambda}}
\lambda\Tr\left\{\left(|n \>\< \nu|
\right)X_{\lambda,i}^{\dagger}\right\} \Tr\left\{\left(|m \>\<
\mu|\right)^{\dagger} X_{\lambda,i}\right\} \nonumber \\
&=&\sum_{\lambda \in \sigma_{|1|}, i=1}^{d_{\lambda}} \lambda
\left(X_{\lambda,i}\right)^{*}_{n \nu} \left(X_{\lambda,i}\right)_{m
\mu}.
\end{eqnarray}
The elements of the dynamical matrix $D_{\Phi_{ass}}$ are
defined by
\begin{equation}
\left(D_{\Phi_{ass}}\right)^{m \mu}_{n
\nu}=\left(\Phi_{ass}\right)^{m n}_{\mu \nu}=\sum_{\lambda \in
\sigma_{|1|}, i=1}^{d_{\lambda}} \lambda
\left(X_{\lambda,i}\right)^{*}_{\mu \nu} \left(X_{\lambda,i}\right)_{m
n}
\end{equation}
so that one obtains the relation
\begin{eqnarray}
\label{dynamical_matrix1}
D_{\Phi_{ass}} &=& \sum_{m,n,\mu,\nu}
\left(D_{\Phi_{ass}}\right)^{m \mu}_{n \nu} |m \mu \>\<n \nu|= \sum_{m,n,\mu,\nu}~\sum_{\lambda \in
\sigma_{|1|}, i=1}^{d_{\lambda}}
\lambda \left(X_{\lambda,i}\right)^{*}_{\mu \nu} \left(X_{\lambda,i}\right)_{m n} |m \mu \>\< n \nu | \nonumber \\
&=& \sum_{m,n,\mu,\nu}~\sum_{\lambda \in
\sigma_{|1|}, i=1}^{d_{\lambda}}
\lambda
\left[\left(X_{\lambda,i}\right)^{}_{m n}|m\>\< n|  \right]
\otimes
\left[\left(X_{\lambda,i}\right)^{*}_{\mu \nu} |\mu\>\<\nu| \right]
= \sum_{\lambda \in
\sigma_{|1|}, i=1}^{d_{\lambda}}
\lambda X_{\lambda,i} \otimes X_{\lambda,i}^{*}.
\end{eqnarray}
Using the identity (\ref{identity_standard_product}) one can rewrite the $d^2 \times d^2$ dynamical matrix as an operator acting
on $d \times d$ matrices according to
\begin{equation}
\label{dynamical_matrix2}
D_{\Phi_{ass}}(A)=\sum_{\lambda \in \sigma_{|1|}, i=1}^{d_{\lambda}}
\lambda X_{\lambda,i} A X_{\lambda,i}^{\dagger},
\end{equation}
where $A$ is an arbitrary $d \times d$ matrix. Expressions
(\ref{dynamical_matrix1}) and (\ref{dynamical_matrix2}) describe the
same dynamical matrix. The first relation describes it as a map acting on reshaped
vectors of length $d^2$ and the second one as a map acting on $d
\times d$ matrices (for details see Appendix
\ref{appendix_dynamical_matrices}). Both expressions are useful to
determine the properties of attractors.

According to Eq.(\ref{dynamical_matrix_properties})
the dynamical matrix (\ref{dynamical_matrix1}) is always hermitian.
Due to proposition \ref{theorem_product of attractors} this property is fulfilled.
Furthermore, the partial trace of the dynamical matrix (\ref{dynamical_matrix1})
over each subsystem yields the identity operator.
For a RUO of the form of Eq.(\ref{unitary_channel})
both properties lead to the condition
\begin{equation}
\sum_{i=1}^{d_1} \Tr\{X_{1,i}^{\dagger}\}X_{1,i}=I.
\end{equation}
so that
the dynamical matrix is positive. With the help of equations (\ref{dynamical_matrix2}) and (\ref{identity_scalar_product})
we find that this positivity is equivalent to the relation
\begin{equation}
\label{inequality_positivity}
\sum_{\lambda \in \sigma_{|1|}, i=1}^{d_{\lambda}}
\lambda \Tr\left\{ A^{\dagger} X_{\lambda,i} A X_{\lambda,i}^{\dagger} \right\} \geq 0
\end{equation}
which has to be fulfilled for an arbitrary $d \times d$ matrix $A$. In view of theorem \ref{comutator} we can thus conclude that the map
\begin{equation}
\Phi_{ass}^{n}(.)=\sum_{\lambda \in
\sigma_{|1|}, i=1}^{d_{\lambda}}
\lambda^n\Tr\{(.)X_{\lambda,i}^{\dagger}\} X_{\lambda,i},
\end{equation}
is a trace-preserving and completely positive unital map for an arbitrary $n \in \Intgrs$.

\section{Discussion and implications}
\label{sec_discussion}
Let us summarize and comment the results obtained so far for the asymptotic behaviour of a
quantum system under a RUO.

First of all, the asymptotic iterative dynamics is determined completely by the attractor set $\set{Atr}(\Phi)$
of a RUO. The Hilbert space $\map B(\hil H)$ can be decomposed as
$\map B(\hil H)=\set{Atr}(\Phi)\oplus
\left(\set{Atr}(\Phi)\right)^{\bot}$ with $\bot$ denoting the orthogonal
complement with respect to $\map B(\hil H)$. Both mutually orthogonal subspaces, i.e.
$\set{Atr}(\Phi)$
and $\alg I_1 = \left(\set{Atr}(\Phi)\right)^{\bot}$ are invariant under the RUO
(\ref{unitary_channel}) and we proved that the component of any initial quantum state
in the subspace
$\alg I_1 $ vanishes after sufficiently large numbers of iterations.
Furthermore, we proved that the vector space of
attractors $\set{Atr}(\Phi)$ is spanned by all elements $X$ of the set $\map B(\hil
H)$ which fulfil
the generalized commutation relations $U_i X = \lambda X U_i$ for
all unitary operators $U_i$ of the decomposition
(\ref{unitary_channel})
and for all eigenvalue $\lambda$ with $|\lambda| = 1$ .

The calculation of the asymptotic iterated dynamics of the random
unitary map (\ref{unitary_channel}) can be divided into four
steps:
\begin{itemize}
\item One determines the set $\sigma_{|1|}$. Usually, this step is
highly nontrivial and depends significantly on the particular
unitary Kraus operators involved. Any additional properties concerning
the structure of the unitary
operators $U_i$ involved, for example, simplify this task considerably.
In particular, the exploitation of symmetries may be useful in this respect.

\item One identifies the set of attractors of the RUO
$\set{Atr}(\Phi)$. This step involves the calculation of all
eigenspaces using the generalized commutation relations
$\set{Ker}(\Phi-\lambda I)=D_{\lambda}$ for all $\lambda \in
\sigma_{|1|}$.

\item One chooses an orthonormal
basis $X_{\lambda,i}$ in each subspace $\set{Ker}(\Phi-\lambda I)$
for $\lambda \in \sigma_{|1|}$.

\item
One
calculates the asymptotic iterated dynamics according to the relation
\begin{eqnarray} \rho(n)&=&
\Phi^n(\rho(0)) (n\gg 1)  = \sum\limits_{\lambda \in \sigma_{|1|},
j=1}^{d_{\lambda}} \lambda^n\Tr\{\rho(0)X_{\lambda,i}^{\dagger}\}
X_{\lambda,i}
\end{eqnarray}
which is valid asymptotically for $n\gg 1$.
\end{itemize}

These general features imply some
important consequences.
Firstly,
the set of attractors $\set{Atr}(\Phi)$ and its corresponding
spectrum is independent of the nonzero probabilities $p_i$ defining
the convex decomposition of the RUO in equation (\ref{unitary_channel}). Thus, two RUOs with the same
unitary operators in their convex decompositions
(\ref{unitary_channel}) have the same attractors space $\set{Atr}(\Phi)$.
The nonzero probabilities $p_i$ determine only how fast an input state
converges to the asymptotic attractor space.

Another simple consequence arises if the ensemble of
random unitary operators defining the RUO $\Phi$ contains the
identity operator $I$ (apart from a global phase). Theorem \ref{comutator}
implies that the only possible eigenvalue of the map $\Phi$ is $\lambda =
1$. Hence from the set of attractors only fixed points can be formed
and the resulting asymptotic dynamics is stationary. Moreover,
assume that the unitary operators $U_i$ are generators of a finite multiplicative
group. As any group contains a unit element all possible eigenvalues
of the RUO $\Phi$
fulfil the relation $\lambda^{n_{\lambda}} = 1$ for some integer
$n_{\lambda} \in \Naturals$. As a consequence the resulting asymptotic dynamics is
periodic.
Such a periodic asymptotic dynamics is also obtained
obtained if the unitary operators
$U_i$ form an irreducible set of operators, i.e. they have no common
nontrivial invariant subspace. This can be proven as follows.
Consider an eigenvalue $\lambda$ of the random unitary operation
(\ref{unitary_channel}) with $|\lambda«|=1$ and its
corresponding eigenvector $X_{\lambda} \neq 0$. Using theorem
\ref{comutator} it can be checked that
$U_i(\set{Ker}(X_{\lambda})) \subset \set{Ker}(X_{\lambda})$ and
$U_i(\set{Ran}(X_{\lambda})) \subset \set{Ran}(X_{\lambda})$ is
fulfilled for all $i \in \hat{m}$. Thus,
$X_{\lambda} \neq 0$ is an invertible
operator. Let
$\alpha \neq 0$ be an eigenvalue of the operator $X_{\lambda}$ and
$u_{\alpha}$ its corresponding eigenvector. From the equation
\begin{equation}
X_{\lambda}U_i u_{\alpha}= \overline{\lambda} U_i X_{\lambda}
u_{\alpha} = \alpha \overline{\lambda} U_i u_{\alpha}
\end{equation}
follows that also $\alpha \overline{\lambda}$ is the eigenvalue of
$X_{\lambda}$ and $U_i u_{\alpha}$ are its
corresponding eigenvectors. Therefore also $\alpha
\overline{\lambda^2}, \alpha \overline{\lambda^3}, ...$ are
eigenvalues of $X_{\lambda}$. Eigenvectors
corresponding to different eigenvalues are linearly independent.
Therefore there is $n \in \Naturals$ such that $\lambda^n=1$.
Moreover, the direct sum of all eigensubspaces corresponding to
eigenvalues $\overline{\lambda^j}\alpha$ ($j \in \{0,1, ...,
n-1\}$) is invariant under all unitary operators $U_i$ and thus
has to be equal to the whole Hilbert space $\hil H$, i.e.
\begin{equation}
\hil H = \bigoplus_{j=0}^{n-1} \set{Ker}(X_{\lambda}-\overline{\lambda^j}\alpha I).
\end{equation}
Therefore,
$X_{\lambda}$ is diagonalizable and can be written in the form
\begin{equation}
\label{projection_form}
X_{\lambda} = \sum_{j=0}^{n-1} \overline{\lambda^j}\alpha (U_i)^j P (U_i^{\dagger})^j,
\end{equation}
where $P$ is the projection on the eigensubspace corresponding to
the eigenvalue $\alpha$ of $X_{\lambda}$ and is determined by
relations of the form
\begin{equation}
\label{commutation_projection}
U_g^iP(U_g^i)^{\dagger}=U_h^iP(U_h^i)^{\dagger}, \hspace{2em} U_h^n P (U_h^n)^{\dagger} = P \neq 0.
\end{equation}
The equation (\ref{commutation_projection}) applies to an
arbitrary pair of unitary operators $U_g$ and $U_h$ in the
decomposition of the random unitary operation
(\ref{unitary_channel}) and their arbitrary i-th power, $i \in
\hat{n}$.

The question remains what happens if the set of unitary operators
$U_i$ is not irreducible. It is shown in the following section that in special cases it may still be possible to
decompose the Hilbert space into so-called minimal
invariant subspaces for which the condition of irreducibility of
unitary operators $U_i$ still holds.

\section{Asymptotic dynamics of a two-qubit CNOT-system}
\label{section_cnots}
In this section we discuss the asymptotic dynamics of the RUO
\begin{equation}
\label{cnot_ ruch}
\Phi(\rho)= p_1 C_1 \rho C_1 + (1-p_1) C_2 \rho
C_2,
\end{equation}
which involves
two controlled-not
(CNOT) operations acting on two qubits.
In the computational basis
$\{|0,0\>,|0,1\>,|1,0\>,|1,1\>\}$ of the two-qubit Hilbert space $\hil
H_2$ the action of these CNOTs is defined by
\begin{equation}
\label{cnots}
C_1 |i,j\> = |i, i \oplus j \>, \hspace{3em} C_2 |i,j\>=|i \oplus
j, j\>
\end{equation}
with
$\oplus$ denoting addition modulo 2. This special RUO of Eq.(\ref{cnot_ ruch}) is a
hermitian operator and therefore its only possible
eigenvalue lying within $\sigma_{|1|}$ are $1$ and $-1$. Let us first
find a decomposition of the two-qubit Hilbert space $\hil H_2$
into subspaces $\hil V_x$, i.e.
\begin{equation}
\hil H_2 = \bigoplus_{x} \hil V_x,
\end{equation}
within each of which
the set of unitary operators $C_1,C_2$ acts irreducibly.
Constructing such a decomposition is equivalent to constructing a
decomposition of the finite multiplicative unitary group $\set C$
generated by $C_1$ and $C_2$. The unitary group $\set C$ is naturally a
unitary representation $I_{\set C}$ of itself.
Therefore, the following considerations are immediate consequences of the standard
theory of representations of finite groups.\cite{Hamermesh}

The unitary group $\set C$ contains six elements divided into
three conjugated classes: $K_1 \equiv$ \{identity element $I$\},
$K_2 \equiv \{C_1C_2,C_2C_1\}$, $K_3 \equiv \{C_1,C_2,C_1C_2C_1\}$. The characters of the representation $I_{\set C}$
corresponding to these classes are $\chi_1=4$, $\chi_2=1$ and $\chi_3=2$. Thus,
there are only three inequivalent irreducible representations of the group $\set C$, say
$D^{\mu}$ ($\mu = \{1,2,3\}$), with dimensions $n_{\mu}$ satisfying the relation
\begin{equation}
\sum_{\mu = 1}^{3} n_{\mu}^2 = 6.
\end{equation}
Hence, there are two one-dimensional and one two-dimensional inequivalent irreducible representations
 of the group $\set C$. The reducible representation $I_{\set C}$ can be expressed in terms of irreducible representations as
 \begin{equation}
 \label{irreducibility decomposition}
 I_C(g) = \sum_{\mu = 1}^{3} a_{\mu} D^{\mu}(g),
 \end{equation}
where $a_{\mu}$ are positive or zero integers and fulfil the relation
\begin{equation}
\label{reducibility_rule}
\sum_{\mu = 1}^3 a_{\mu}^2 = \frac{1}{g} \sum_{i = 1}^{3} g_i = |\chi_{i}|^2= 5
\end{equation}
with
$g$ and $g_i$ denoting the number group elements and
the number of elements of the conjugated class $K_i$, respectively. The
only possibility to satisfy the dimensionality of the representation
$I_{\set C}$ and equation (\ref{reducibility_rule}) is the
solution: $a_1 = 1$ for the two-dimensional irreducible
representation, $a_2 = 2$ for the one-dimensional irreducible
representation, the second one-dimensional irreducible
representation cannot be involved in the decomposition, i.e.
$a_3=0$. Two one-dimensional representations contained in the
irreducible decompositions (\ref{irreducibility decomposition}) mean
that there are just two common eigenvectors for the unitary group
$\set C$ and thus common eigenvectors of operators $C_1$ and $C_2$.
>From the definition (\ref{cnots}) it is clear that these
eigenvectors are $e_1 = |00\>$ and
$e_4 = (|01\>+|10\>+|11\>)/\sqrt{3}$. Subsequently, we know that
the minimal invariant subspaces of operators $C_1$ and $C_2$ are:
$V_1 = \span(e_1)$, $V_2 =
\span\left(e_2=\frac{1}{\sqrt{2}}(|01\>-|10\>),e_3=\frac{1}{\sqrt{6}}(|01\>+|10\>-2|11\>)\right)$
and  $V_3 =
\span(e_4)$. If we
denote the restriction of the operator $C_i$ to the subspace $V_x$
as $C_i^{(x)}$,
in the orthonormal basis system
$\{e_i\}_{i=1}^{4}$
the operators $C_1$ and $C_2$ correspond to the matrices
\begin{equation}
C_1=\left( \begin{array}{ccc} C_1^{(1)} & 0 & 0 \\ 0& C_1^{(2)} & 0 \\ 0& 0& C_1^{(3)}\end{array}
\right)= \left( \begin{array}{cccc} 1 & 0 & 0 & 0 \\ 0& \frac{1}{2} & \frac{\sqrt{3}}{2}&
0 \\ 0& \frac{\sqrt{3}}{2}& -\frac{1}{2} & 0 \\ 0&0&0&1  \end{array}\right), \hspace{2em}
C_2 = \left( \begin{array}{ccc} C_2^{(1)} & 0 & 0 \\ 0& C_2^{(2)} & 0 \\ 0& 0& C_2^{(3)}\end{array}
\right)= \left( \begin{array}{cccc} 1 & 0 & 0 & 0 \\ 0& \frac{1}{2} & -\frac{\sqrt{3}}{2}& 0 \\
0& -\frac{\sqrt{3}}{2}& -\frac{1}{2} & 0 \\ 0&0&0&1  \end{array}
\right).
\end{equation}
Writing the general commutation relations (\ref{comutator}) in the
block structure form we obtain for $i \in \{1,2\}$
\begin{equation}
\label{block_commutator}
C_i^{(m)} X^{(mn)} = \lambda X^{(mn)} C_i^{(n)}
\end{equation}
with the $1\times1$- matrices
$X^{(11)},X^{(13)},X^{(31)}, X^{(33)}$, with the  $2\times2$ matrix
$X^{(22)}$, with the
$1\times2$-matrices
$X^{(12)}$ and $X^{(32)}$, and with the
$2\times1$ matrices
$X^{(21)}$ and $X^{(23)}$.
Using equations (\ref{block_commutator}) one can check
\begin{equation}
C_i^{(n)}(\set{Ker}(X^{(mn)})) \subset \set{Ker}(X^{(mn)}), \hspace{2em}
C_i^{(m)}(\set{Ran}(X^{(mn)})) \subset \set{Ran}(X^{(mn)})
\end{equation}
and thus
$X^{(mn)}$ is either the zero operator or an invertible
operator. Hence, $X^{(12)},X^{(32)},X^{(21)},X^{(23)}$ are inevitably
zero matrices.

Now, assume the case $\lambda=1$. A simple evaluation of
equation (\ref{block_commutator}) leads to the relations $X^{(11)}=a$,
$X^{(33)}=b$, $X^{(13)}=c$, and  $X^{(31)}=d$ ($a,b,c,d \in \Cmplx$). The
remaining matrix block $X^{(22)}$ has to commute with the
irreducible set of $2\times2$ matrices $C_i^{(2)}$ ($i \in \{1,2\}$)
and has to be equal to a multiple of the identity matrix
$X^{(22)}=eI$ ($e \in \Cmplx$). The eigenspace of the random unitary
operation (\ref{cnot_ ruch}) corresponding to eigenvalue $1$ is
five-dimensional and the most general eigenvector reads
\begin{equation}
X_1=\left( \begin{array}{cccc} a & 0 & 0 & c \\ 0& e & 0& 0 \\ 0&
0& e & 0\\ d&0&0&b  \end{array}\right).
\end{equation}
The solution of Eq.(\ref{block_commutator}) with $\lambda=-1$ yields
$X^{(11)}=X^{(13)}=X^{(31)}=X^{(33)}=0$. The last matrix block $X^{(22)}$ is
determined by anticommutation relations with the irreducible set of operators
$C_i^{(2)}$ ($i \in \{1,2\}$), i.e.
\begin{equation}
C_i^{(2)}X^{(22)}=-X^{(22)}C_i^{(2)}X^{(22)}.
\end{equation}
>From the discussion in section \ref{sec_discussion} and the Eq. (\ref{projection_form})
follows that $X^{(22)}$ is either the zero operator or
\begin{equation}
\label{projection_form_cnot}
X^{(22)}=f(P-C_i^{(2)}PC_i^{(2)})  \hspace{2em} f \in \Cmplx
\end{equation}
with
the projection $P$ being determined by the equation
\begin{equation}
C_1^{(2)}C_2^{(2)}P=PC_1^{(2)}C_2^{(2)}.
\end{equation}
Hence, the projection operator $P$ is diagonal in the eigenbasis of the operator $C_1^{(2)}C_2^{(2)}$.
Using Eq.(\ref{projection_form_cnot}) the most general form of the matrix block
$X^{(22)}$ corresponding to eigenvalue $-1$ reads
\begin{equation}
X^{(22)}=\left( \begin{array}{cc} 0 & f \\ -f & 0 \end{array} \right), \hspace{2em} f \in \Cmplx.
\end{equation}
Thus,
the eigenspace of
the random unitary operation (\ref{cnot_ ruch}) corresponding to
eigenvalue $-1$ is one-dimensional and the general eigenvector
reads
\begin{equation}
X_{-1}=\left( \begin{array}{cccc} 0 & 0 & 0 & 0 \\ 0& 0 & f& 0 \\ 0&
-f& 0 & 0\\ 0&0&0&0  \end{array}\right).
\end{equation}
Therefore, in the computational basis
the attractor space is spanned by the matrices
\begin{eqnarray}
X_{1,1}&=&\left( \begin{array}{cccc} 1 & 0 & 0 & 0 \\  0 & 0 & 0 & 0 \\  0 & 0 & 0 & 0 \\  0 & 0 & 0 & 0 \end{array} \right), \hspace{1em} X_{1,2}=\frac{1}{\sqrt{6}}\left( \begin{array}{cccc} 0 & 0 & 0 & 0 \\  0 & 0 & 1 & 1 \\  0 & 1 & 0 & 1 \\  0 & 1 & 1 & 0 \end{array} \right), \hspace{1em} X_{1,3}=\frac{1}{\sqrt{3}}\left( \begin{array}{cccc} 0 & 1 & 1 & 1 \\  0 & 0 & 0 & 0 \\  0 & 0 & 0 & 0 \\  0 & 0 & 0 & 0 \end{array} \right), \nonumber \\  X_{1,4}&=& \frac{1}{\sqrt{3}}\left( \begin{array}{cccc} 0 & 0 & 0 & 0 \\  1 & 0 & 0 & 0 \\  1 & 0 & 0 & 0 \\ 1 & 0 & 0 & 0 \end{array} \right), \hspace{1em} X_{1,5}=\frac{1}{\sqrt{3}}\left( \begin{array}{cccc} 0 & 0 & 0 & 0 \\  0 & 1 & 0 & 0 \\  0 & 0 &1 & 0 \\  0 & 0 & 0 & 1 \end{array} \right), \hspace{1em} X_{-1,1}=\frac{1}{\sqrt{6}}\left( \begin{array}{cccc} 0 & 0 & 0 & 0 \\  0 & 0 & -1 & 1 \\  0 & 1 & 0 & -1 \\  0 &-1 & 1 & 0 \end{array} \right).
\end{eqnarray}
In this notation
the first index refers to the eigenvalues of the RUO (\ref{cnot_ ruch}) and the second index runs through
the basis states of the corresponding eigenspaces.

Finally, consider the most general two-qubit input density matrix
\begin{equation} \rho(0) = \left( \begin{array}{cccc} a_{11} &
a_{12} & a_{13}& a_{14} \\ a_{12}^* & a_{22} & a_{23}& a_{24} \\
a_{13}^* & a_{23}^* & a_{33}& a_{34} \\ a_{14}^* & a_{24}^* &
a_{34}^*& a_{44}\end{array} \right). \nonumber
\end{equation}
Thus,
theorem \ref{theorem:final state} implies that the asymptotic dynamics under the RUO (\ref{cnot_ ruch}) is periodic with period two
and is determined by the relations
\begin{equation}
\lim_{n \rightarrow + \infty} \rho(2n) = \left( \begin{array}{cccc} a & c & c& c
\\ c^* & b & d& d^* \\ c^* & d^* &
b& d \\ c^* & d & d^*& b\end{array} \right), \hspace{1em}
\lim_{n \rightarrow + \infty} \rho(2n+1) = \left( \begin{array}{cccc} a & c & c& c
\\ c^* & b & d^*& d \\ c^* & d &
b& d^* \\ c^* & d^* & d& b\end{array} \right)
\end{equation}
with
$a=a_{11}$, $b=\frac{1}{3}(a_{22}+a_{33}+a_{44})$,
$c=\frac{1}{3}(a_{12}+a_{13}+a_{14})$,
$d=\frac{1}{3}(a_{23}+a_{34}+a_{24}^*)$. This two-qubit CNOT network
is one of the simplest examples of a network allowing for
oscillatory asymptotic dynamics.

\section{Conclusions and outlook}
We studied general properties of random unitary operations and
presented several theorems allowing to determine the asymptotic long
time dynamics. Thereby, a central result is the structure theorem
which states
that the asymptotic states are located completely inside the vector space
spanned by a typically small set of attractors.
The
form of these asymptotic quantum states depends on this attractor space and on
the choice of the
initial state but is independent of
the actual values of the probabilities with which the unitary transformations are applied. However,
these
probabilities affect the rate of the convergence
towards the asymptotic quantum state.

It should be stressed that the asymptotic dynamics need not result in a
stationary state.
Thus, in
contrast to thermalization
the asymptotic dynamics might also be periodic as
illustrated by the example of two qubits interacting by random C-NOT operations.
Even an aperiodic non-stationary asymptotic dynamics is possible.

The obtained results rise several additional questions.
First of all, it is not yet clear
what determines the convergence rate of a quantum system
towards its asymptotic dynamics. Numerical studies suggest that
in many cases this convergence has an exponential character which depends on the probabilities with
which the unitary operations are applied.
Preliminary results also suggest
that at least in the case of many-qubit networks involving controlled-not operations
the topology of the
network is related to
the set of attractors.

Finally, it should also be mentioned  that our results might have
applications for quantum operations which involve an averaging procedure over
a group, such as twirling operations.
Our results might allow to
choose efficiently the minimal set of unitary transforms leading to
a particular asymptotic state.
In addition,
we expect that the theory presented might also contribute to
other related problems concerning the determination of eigenvectors of
random unitary maps \cite{Pasczio2004} or their application
in purification protocols.\cite{Toth}

\section*{Acknowledgements}
Financial support by the Czech Ministry of Education, by MSM 6840770039, M\v SMT LC 06002, by the AvH Foundation, and by  the DAAD 
is gratefully acknowledged.
\appendix
\section{Jordan canonical form}
\label{app:jordan}
Let us recall the
definition and properties of the Jordan canonical form of square
matrices. Consider a complex square matrix $A=(A_{ij})_{i,j=1}^n$
of size $n \times n$ ($A\in \Cmplx^{n \times n}$). It is
similar to a block diagonal matrix
\begin{equation}
\label{jordan_form} A=\left( \begin{array}{cccc} J_1 & & & \\ & J_2 & & \\
& & \ddots & \\ & & & J_p
\end{array}\right),
\end{equation}
in which
each Jordan block $J_i$ ($i \in \hat{p} := \{1,...,p\}$) is given by
\begin{equation}
J_{i}= \left( \begin{array}{cccc} \lambda_i & 1& & \\ & \lambda_i &\ddots & \\
& & \ddots & 1\\ & & & \lambda_i
\end{array}\right).
\end{equation}
Thus, there is an invertible matrix $P \in \Cmplx^{n
\times n}$ such that $A=PJP^{-1}$ or equivalently there is a
Jordan basis $x_{i,\alpha} \in \Cmplx^n$ ($i \in \hat{p}$, $\alpha
\in \left\{1,2, \hdots,\dim(J_i)\right\}$) in which the linear
map corresponding to the matrix $A$ has the diagonal form
(\ref{jordan_form}). In general, this basis is non-orthogonal and the
vectors $x_{i,\alpha}$ ($\alpha \in \left\{1,2,
\hdots,\dim(J_i)\right\}$) form the basis of the Jordan block $J_i$
which corresponds to the eigenvalue $\lambda_i$ of the matrix $A$.

The geometric multiplicity of the eigenvalue $\lambda_i$ is the
number of Jordan blocks corresponding to $\lambda_i$ and the sum
of the sizes of all Jordan blocks corresponding to an eigenvalue
$\lambda_i$ is its algebraic multiplicity. Therefore, the matrix
$A$ is diagonalizable if and only if all Jordan blocks are one dimensional.
In all other cases any Jordan block, say $J_i$,
with dimension $s>1$ gives rise to a Jordan chain. This means
that there is a so-called lead vector or generator, say
$x_{i,\dim(J_i)}$, which is a generalized
eigenvector, i.e. $(A-\lambda_i I)^s x_{i,\dim(J_i)}=0$. The vector
$x_{i,1} = (A - \lambda_i I)^{s-1}x_{i,\dim(J_i)}$ is an
eigenvector corresponding to the eigenvalue $\lambda_i$. In
general, the vector $x_{i,j}$ is the image of the vector
$x_{i,j+1}$ under the linear map $A-\lambda_i I$. In this sense
all vectors $x_{i,\alpha}$ ($\alpha \in \left\{2,3,
\hdots,\dim(J_i)\right\}$) are generalized eigenvectors of the matrix
$A$.

Therefore, for every square matrix $A$ there exists a basis
consisting only of eigenvectors and generalized eigenvectors of
the matrix $A$ in which the matrix $A$ can be put in Jordan normal
form (\ref{jordan_form}).

\section{Dynamical matrices}
\label{appendix_dynamical_matrices} Let us summarize the concept of
\emph{dynamical matrices} which is useful to understand problems
related to complete positivity of maps. We just recall its
definition and present a short summary of characteristic properties needed in the
main body of our text. Detailed proofs are given in Ref. [7], for example.

Assume that $A$ is an operator acting on a $d$-dimensional Hilbert space $\hil H_d$. Hence $A_{ij}$ ($i,j \in \hat{d}:=\{1,...,d\}$)
are its matrix elements with respect to a given orthonormal basis. It is convenient to interpret
a $d\times d$-matrix $(A)_{ij}$ as a vector $\vec A=(A_{m\mu}) \in \hil H_{d^2}$ of the length $d^2$
\begin{equation}
\label{vector_notation}
\vec A=(A_{11},A_{12},\ldots,A_{1d},A_{21},A_{22},\ldots,A_{2d},\ldots,A_{d1},A_{d2},\ldots,A_{dd}).
\end{equation}
One can check that two $d\times d$ matrices $A$ and $B$ fulfil
\begin{equation}
\label{identity_scalar_product}
\<A,B\> \equiv \Tr\{A^{\dagger}B\}=\vec A^{*} \vec B= \<\vec A,\vec B\>.
\end{equation}
The vector $\vec A$ of the length $d^2$ may be linearly transformed into the vector $\vec A^{'}=C \vec A$ by a matrix $C$ of size $d^2 \times d^2$ whose matrix elements may be denoted by $C_{kk^{'}}$ with $k,k^{'} = 1,\ldots,d^2$.
In addition, it is also convenient to use a four index notation $C^{m\mu}_{n \nu}$
with respect to a two index notation of vectors (\ref{vector_notation})
with $m,n,\mu,\nu = 1,\ldots,d^2$.
The matrix $C$ may represent an operator acting in a composite Hilbert space $\hil H=\hil H_d \otimes \hil H_d$.
The tensor product of any two orthonormal basis systems in both factors provides a basis in $\hil H$ so that we obtain
\begin{equation}
C^{m\mu}_{n \nu}=\<e_m \otimes f_{\mu}|C|e_n \otimes f_{\nu}\>
\end{equation}
with
Latin indices referring to the first subsystem, $\hil H_A=\hil H_d$, and Greek indices to the
second subsystem, $\hil H_B=\hil H_d$.
The operation of partial trace over the second or first subsystem produces the $d \times d$ matrices $C^A \equiv \Tr_B C$ or
 $C^B \equiv \Tr_A C$, respectively, i.e.
\begin{equation}
C^A_{mn}=\sum_{\mu=1}^{d^2}~C^{m \mu}_{n \mu}, \hspace{3em} {\rm and} \hspace{1em} C^B_{\mu \nu}=\sum_{m=1}^{d^2}~C^{m \mu}_{m \nu}.
\end{equation}
If $C=A \otimes B$, then $C^{m \mu}_{n \nu}=A_{mn}B_{\mu \nu}$.
The standard product of three matrices can be rewritten in the following useful form
\begin{equation}
\label{identity_standard_product}
ABC= \Phi \vec B \hspace{3em} {\rm with} \hspace{2em} \Phi=A \otimes C^{T}.
\end{equation}
With the help of
identity (\ref{identity_standard_product}) we can rewrite the RUO (\ref{unitary_channel}) in the form
\begin{equation}
\Phi=\sum_{i=1}^{m}p_i U_i \otimes U_i^{*}.
\end{equation}
Here, the RUO $\Phi$ is not understood as a map acting on the $d \times d$-dimensional matrix space
but as a map acting on the vector space of the dimension $d^2$.

Let $\Phi$ be a completely positive trace-preserving map mapping an arbitrary $d \times d$ density matrix $\rho\in {\cal B}(\hil H_d)$
of a $d$-dimensional Hilbert space $\hil H_d$ on a density matrix $\rho^{'}\in {\cal B}(\hil H_d)$, i.e.
\begin{equation}
\rho^{'}=\Phi \rho  \hspace{2em} {\rm or} \hspace{2em} \rho^{'}_{m \mu}=\sum_{n,\nu=1}^{d^2}~\Phi^{m \mu}_{n \nu} \rho_{n \nu}.
\end{equation}
The meaning of complete positivity becomes rather transparent if we reshuffle $\Phi$ and define the \emph{dynamical matrix} $D_{\Phi}$
\begin{equation}
(D_{\Phi})^{mn}_{\mu \nu}= \Phi^{m \mu}_{n \nu}.
\end{equation}
The dynamical matrix $D_{\Phi}$ uniquely determines the map $\Phi$ and has the following properties
\begin{equation}
\label{dynamical_matrix_properties}
\begin{array}{rccl}
{\rm (i)}\hspace{2em} & \rho^{'}=(\rho^{'})^{\dagger}& \Leftrightarrow & D_{\Phi}=D_{\Phi}^{\dagger} \\
{\rm (ii)} \hspace{2em} & \Tr \rho^{'}=1 & \Leftrightarrow & \Tr_A D_{\Phi}= I \\
{\rm (iii)} \hspace{2em} & \Phi(I)=I \hspace{1em} {\rm (unital)}& \Leftrightarrow & \Tr_B D_{\Phi}=I \\
{\rm (iv)} \hspace{2em}& \Phi \hspace{1em} {\rm is \hspace{0.5em} CP \hspace{0.5em} map} & \Leftrightarrow & D_{\Phi} \hspace{1em} {\rm is \hspace{0.5em} positive}.
\end{array}
\end{equation}

\end{document}